\documentclass[pre,twocolumn,a4paper,showpacs]{revtex4-1}
\usepackage{amssymb}
\usepackage{wasysym}
\usepackage{graphicx}
\usepackage{epsfig}
\usepackage{subfigure}
\begin{document}

\title{Influence  of  network topology on  cooperative problem-solving   systems}

\author{Jos\'e F. Fontanari}

\affiliation{Instituto de F\'{\i}sica de S\~ao Carlos,
  Universidade de S\~ao Paulo,
  Caixa Postal 369, 13560-970 S\~ao Carlos, S\~ao Paulo, Brazil}
  
  \author{Francisco A. Rodrigues}

\affiliation{Instituto de Ci\^encias Matem\'aticas e de Computa\c{c}\~ao,
              Universidade de S\~ao Paulo,
             Caixa Postal 668, 13560-970 S\~ao Carlos, S\~ao Paulo, Brazil}

\begin{abstract}
The idea of a collective intelligence behind the  complex natural structures built by organisms suggests that the organization of social networks is selected so as to 
optimize problem-solving competence at the  group-level. Here we study the influence of the social network topology on the  performance of a group of agents
whose task is to locate the global maxima of  NK fitness landscapes. Agents cooperate by broadcasting  messages  informing on their fitness and
use this information to imitate  the fittest agent in their influence networks.  In the case those messages  convey accurate  information on the 
proximity of the solution (i.e., for smooth fitness landscapes) we find that high connectivity as well as centralization boost   the group  performance. For rugged landscapes, however, these characteristics are beneficial for small groups  only. For large groups, it is advantageous to slow down the information transmission through the
network to avoid local maximum traps. Long-range links and modularity have marginal effects on the performance of the group, except for a very narrow region of the model parameters.
\end{abstract}

\maketitle

\section{Introduction}\label{sec:intro}

The benefits of cooperative work  have long been explored by nature as revealed by  the collective structures built by termites and slime molds. These structures may be thought of as the organisms'  solutions to the problems that endanger their existence~\cite{Bloom_01,Queller_09}. Within this (admittedly controversial) perspective, competence in problem solving should be viewed as a selection pressure  on the size and organization of groups of gregarious animals,  in addition to the more traditional pressures such as defense against predation and foraging success   \cite{Wilson_75}.  In fact, given the ubiquity of network optimization in biology \cite{Barabasi_04}, one would expect that the size and the topology of social networks are  also selected for optimization of species adaptation and surveillance~\cite{Derex_13,Pasquaretta_14}.
Social networks are  considered optimal if they improve communication and decision making at the group level, while keeping a  minimum number of (costly) connections between the individuals \cite{Pasquaretta_14}. 

The information in social networks flows between individuals via social contacts and, in the problem-solving  context, the relevant process is imitative learning or, more generally, social learning \cite{Heyes_94}, as  expressed in this quote by Bloom ``Imitative learning acts like a synapse, allowing information to leap the gap from one creature to another'' \cite{Bloom_01}. In fact, social learning has inspired the planning of a variety of optimization heuristics, such as the particle swarm optimization algorithm~\cite{Bonabeau_99} and the adaptive culture heuristic~\cite{Kennedy_98,Fontanari_10}.  Although   there has been some progress on the  understanding of  the factors that make cooperative group work effective \cite{Huberman_90,Clearwater_91,Hong_04,Rendell_10},  a workable minimal model to study  group problem-solving  via imitative learning was proposed only very recently~\cite{Fontanari_14,Fontanari_15}.  Our goal here is to investigate how the social network properties are related to efficiency at the group level within that minimal model framework.  By efficiency we mean that the time to solve a problem must scale superlinearly with the number of individuals or resources employed in the task.

The agent-based model we consider consists of a group of $L$ agents that can perform individual trial-and-test searches to explore a fitness landscape and, most importantly,  can imitate a model agent -- the best performing agent in their influence network at the trial -- with a certain probability $p$.  Thus the model exhibits the two critical ingredients of a collective brain: imitative learning and a dynamic hierarchy among the agents \cite{Bloom_01}. It is 
the exchange of messages   between agents  informing each other on their partial success (i.e., their fitness at the current trial)  towards the completion of the task  that characterizes the model as a  distributed cooperative problem-solving system \cite{Huberman_90}.

The task of the agents is to find the global maxima of smooth and rugged fitness landscapes created by the NK model  and
the performance or efficiency of the group is measured by the number of trials required to find those maxima. We assume that the links between agents are costless. Differently from  a previous study that  considered only fully connected networks (i.e., each agent influences and is influenced by all the other agents in the group)  \cite{Fontanari_15}, here we focus on the structure of the social network that connects the agents within the group. In particular, we consider the effect of varying the connectivity (or coordination number) of one-dimensional lattices as well as of varying the structure of the network  while keeping the average connectivity of the network fixed. 

We find that in both smooth and rugged landscapes the best group organization comprises an optimal number  $L^*$ of  fully connected agents. For smooth landscapes, decrease of the connectivity and hence of the  speed of information transmission is always detrimental  to the performance of the group, regardless of the group size $L$.  For rugged landscapes, however, this decrease
can be greatly beneficial if $L > L^*$, but it is slightly detrimental  if $L < L^*$. In the case the number of agents and links are kept fixed, we find 
that centralization (i.e., the existence of agents with a large number of connections) is, in general, advantageous for small size groups. In addition,
centralization in large groups leads to catastrophic performances in rugged landscapes. We find that the introduction of long-range connections has no effect on the performance of the group, except when the parameters are set close to the region where the imitative search fails to find the global maximum. In that case those links
greatly harm the group performance. It is interesting to notice that this behavior is different from other dynamical processes taking place in networks. For instance, synchronization is strongly influenced by the decrease of the average shortest distances in the network~\cite{Barahona_02}.

The rest of the paper is organized as follows. In Section \ref{sec:NK}  we present a brief description of the NK model of  
rugged fitness landscapes.  In Section \ref{sec:model}  we describe in some detail  the imitative learning search strategy and introduce a convenient measure of the performance of the search for the global maximum on NK landscapes. In Section \ref{sec:latt} we study the performance of the imitative search
in the case the imitation network  is modeled by a regular one-dimensional lattice with coordination number $M$ and in Section \ref{sec:comp} in the case  it is modeled by  more complex network topologies, namely, scale-free, small-world and community networks.
 Finally, Section \ref{sec:conc} is reserved to our concluding remarks.

\section{NK Model of Rugged Fitness Landscapes}\label{sec:NK}

The NK model was introduced by Kauffman \cite{Kauffman_87} to model the adaptive evolution process as walks on rugged fitness landscapes. The main advantage of the NK model of rugged fitness landscapes is the possibility of tuning  the ruggedness of the landscape by changing the two integer parameters  of the model, namely, $N$ and $K$.
The NK landscape is defined in the space of binary strings of length $N$. This parameter determines the size of the solution space, $2^N$.  The other parameter  $K =0, \ldots, N-1$ influences the number of local maxima on the landscape. In particular,
for $K=0$ the corresponding (smooth) landscape has one single maximum. For $K=N-1$, the (uncorrelated) landscape  has on the average  $2^N/\left ( N + 1 \right)$ maxima
with respect to single bit flips \cite{Kaul_06}. 

\begin{figure}
\centering
\includegraphics[width=1\columnwidth]{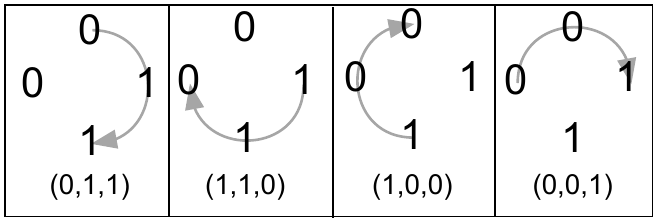}
\caption{Example of the computation of the fitness for the  string $\mathbf{x} = (0,1,1,0)$ with $N=4$ and  $K=2$. There are four components, as indicated in the figure. For instance, if $\phi(0,1,1) = 0.3$, $\phi(1,1,0) = 0,2$, $\phi (1,0,0) = 0.5$ and $\phi(0,0,1) = 0.1$, then $\Phi \left ( \mathbf{x}  \right ) = (1/4)(\phi_0+\phi_1+\phi_2+\phi_3) = 0.275$. This procedure is performed for each of the $2^N$ possible strings that define the landscape.} 
\label{fig:nk}
\end{figure}

The solution space of the NK landscape consists of the $2^N$ distinct binary strings of length $N$, which we denote by $\mathbf{x} = \left ( x_1, x_2, \ldots,x_N \right )$ with
$x_i = 0,1$. To each string $\mathbf{x}$ we associate a fitness value $\Phi \left ( \mathbf{x}  \right ) $ which  is an average  of the contributions from each 
component $i$ in the string, i.e.,
\begin{equation}
\Phi \left ( \mathbf{x}  \right ) = \frac{1}{N} \sum_{i=1}^N \phi_i \left (  \mathbf{x}  \right ) ,
\end{equation}
where $ \phi_i$ is the contribution of component $i$ to the  fitness of string $ \mathbf{x} $. It is assumed that $ \phi_i$ depends on the state $x_i$  as well as on the states of the $K$ right neighbors of $i$, i.e., $\phi_i = \phi_i \left ( x_i, x_{i+1}, \ldots, x_{i+K} \right )$ with the arithmetic in the subscripts done modulo $N$. We notice that $K$ measures the degree of interaction (epistasis) among the components of the strings. It is assumed, in addition, that  the functions $\phi_i$ are $N$ distinct real-valued functions on $\left \{ 0,1 \right \}^{K+1}$. As usual,  we assign to each $ \phi_i$ a uniformly distributed random number  in the unit interval \cite{Kauffman_87}. Hence $\Phi \in \left ( 0, 1 \right )$ has a unique global maximum. Figure \ref{fig:nk} illustrates the calculation of $\Phi \left ( \mathbf{x}  \right )$.

%----------------------------------------------------------------------------------------------------------------
%------------------------------------------------------------------------------------------------------------
%

For $K=0$ there are no local maxima and the sole maximum of $\Phi$ is easily located by picking for each component $i$ the state $x_i = 0$ if  $\phi_i \left ( 0 \right ) >  \phi_i \left ( 1 \right )$ or the state  $x_i = 1$, otherwise. However, for $K>0$ finding the global maximum of the NK model is a NP-complete problem \cite{Solow_00}, which  means that the time required to solve the problem using any currently known deterministic algorithm increases exponentially fast with the length $N$ of the strings \cite{Garey_79}.
The increase of the parameter $K$ from $0$ to $N-1$  decreases the correlation between the fitness of neighboring configurations 
(i.e., configurations that differ at a single component) in the solution space. For $K=N-1$, those fitness values are  uncorrelated
so the NK model reduces to the Random Energy model \cite{Derrida_81,David_09}.  The simplest way to see this is to consider  two neighboring configurations, say
$\mathbf{x}^a = \left ( 0, 0, \ldots, 0 \right )$ and $\mathbf{x}^b = \left ( 1, 0, \ldots, 0 \right )$, and  to calculate explicitly the correlation between their fitnesses, which yields $\mbox{corr} \left ( \Phi \left ( \mathbf{x}^q  \right ),   \Phi \left ( \mathbf{x}^b  \right ) \right ) = 1 - \left ( K+1 \right )/N $, as expected.

Since the functions  $ \phi_i$ are random, the ruggedness measures (e.g., the number of local maxima) of a particular realization of a NK landscape are not fixed by the parameters $N$ and $K$. In fact, those measures  can vary considerably between landscapes characterized by the same values of $N$ and $K>0$ \cite{Kauffman_87}, which implies that the  performance of any search  heuristic based on the  local correlations of the fitness landscape  depends on the particular realization of the landscape. Hence  in order to highlight the role of the parameters  that are relevant to imitative learning -- the group size $L$, the network topology and average connectivity $M$,  and the imitation probability $p$ --
here we compare the performance  of  groups characterized by different sets of those parameters for the same realization of the NK fitness landscape. In particular, we consider two types of landscape: a smooth landscape with  $N=12$ and $K=0$  and  a rugged landscape with $N=12$ and $K=4$. The
effects of averaging over different landscape realizations is addressed briefly in Section \ref{sec:comp}.

\section{The model}\label{sec:model}

\begin{figure}
\centering
\includegraphics[width=0.9\columnwidth]{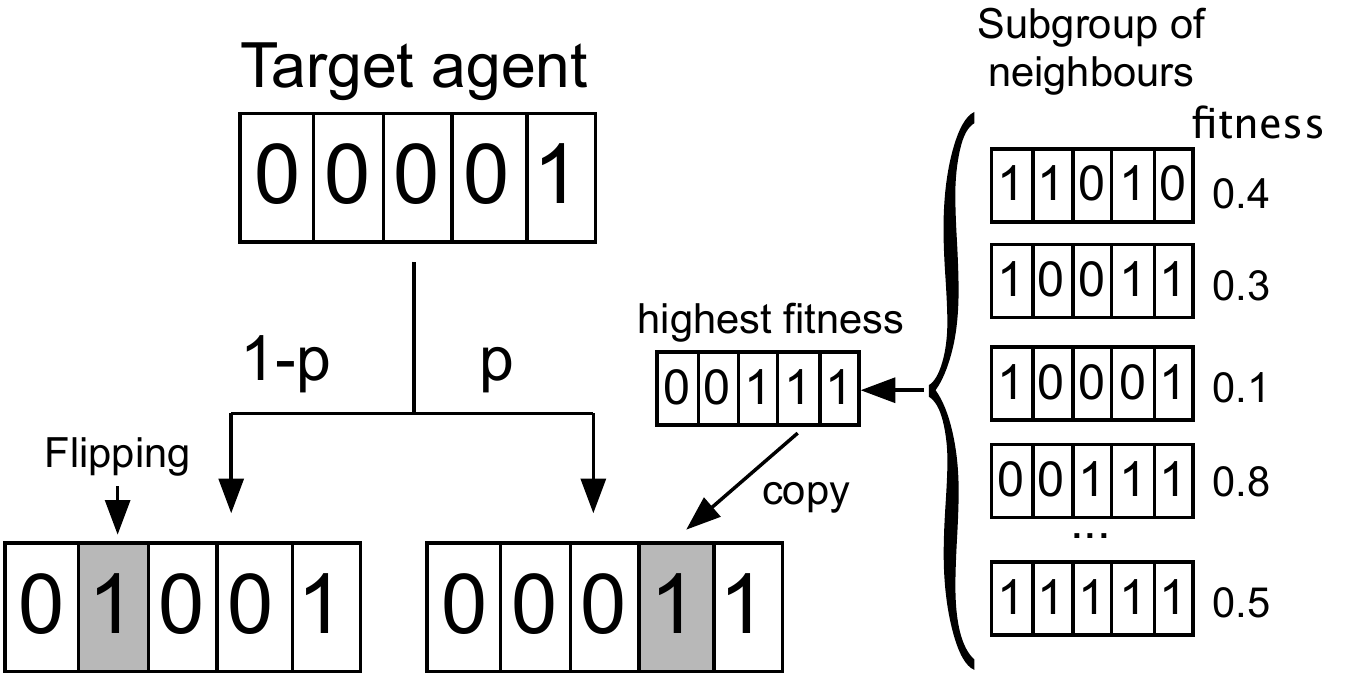}
\caption{Illustration of the update rules of the model. At any step, a target agent is selected and its binary string is changed either by flipping one randomly  chosen bit  or by copying a bit from the highest fitness (model) string. The former process occurs with probability $1-p$, whereas the latter with probability $p$. Notice that when the imitation operation is selected, the updated string becomes closer to the model string.}
\label{fig:model}
\end{figure}

We consider a  group  composed of $L$  agents.  Each agent operates in an initial binary string drawn at  random with equal probability for the bits $0$ and $1$. Henceforth we  use here the terms agent and string interchangeably. At any trial $t$, a target agent can choose  between two distinct processes to operate  on the strings. The first process,
which happens with probability $1-p$,  is the elementary move in the solution space that consists of picking a  bit  $i=1, \dots, N$ at random with equal probability and then flipping it. Through the repeated application of this operation,  the agent can produce all  the $2^N$  binary strings starting from any arbitrary string. The second process, which  happens with probability $p$, is the  imitation  of a model string. We choose the model string as the highest fitness string at  trial $t$  among the (fixed) subgroup of agents that  can influence the target agent. The string to be updated (target string)  is compared with the model string and the different bits are singled out.  Then the agent  selects at random one of the distinct bits and flips it so that this bit is now the same in both the  target and the model  strings. Hence, as a result of the imitation process the target string becomes more similar to the model string. Figure~\ref{fig:model} illustrates the steps involved in the model.

%----------------------------------------------------------------------------------------------------------------
%------------------------------------------------------------------------------------------------------------

The parameter $p \in \left [0,1 \right ]$ is the imitation probability. The case $p=0$ corresponds to the baseline situation in which  the $L$ agents explore the solution space independently.  The imitation procedure described above is motivated by the mechanism used to simulate the influence of an external media  \cite{Shibanai _01,Peres_11} in  the celebrated agent-based model proposed by  Axelrod to study the process of culture dissemination  \cite{Axelrod_97,Barbosa_09}. We notice that in the case the target string is identical to the model string, and this situation is not uncommon since the imitation process reduces the diversity of the group, the agent executes the elementary move with probability one.

The collective search ends when one of the agents finds the global maximum and we denote by $t^*$ the number of  trials made by the agent that found the solution.  Since the agents operate in parallel and the search halts simultaneously for all agents,  $t^*$ stands also for the number of trials made by any one of the $L$ agents. In other words, the trial number $t$ is  incremented by one unit when the $L$ agents have executed  one of the two operations on its associated string. We notice that except for the case  $p=0$, the update of the $L$ agents is not strictly a parallel process since the model strings may change several times within a given trial. Nonetheless, since in a single trial all agents are updated, the total number of agent updates at trial $t$ is given by the product $Lt$.

Here we measure the efficiency of the  search by the total number of agent updates necessary to find the global maximum (i.e., $Lt^*$), which is essentially  the computational cost of the search.  Since the typical number of trials to success $t^*$ scales with the size of the solution space $2^N$, it is convenient to present the results in terms of the rescaled computational cost,  defined as $C \equiv L t^*/2^N$. 
In fact, in the case of the independent search it can be shown that the mean rescaled computational cost is given by \cite{Fontanari_15}
\begin{equation}\label{Cind}
\langle C \rangle = \frac{L}{ 2^N \left [ 1 - \left ( \lambda_N \right)^L \right ]},
\end{equation}
where  $\lambda_N$ is the second largest eigenvalue of a tridiagonal stochastic matrix $\mathbf{T}$ whose elements are 
$T_{ij} = \left ( 1-j/N \right ) \delta_{i, j+1} + j/N \delta_{i, j-1}$ for $j=1,\ldots,N-1$, $T_{i0} = \delta_{i,1}$, and  $T_{iN} = \delta_{i,N}$, where $\delta_{i, j} $ is the Kronecker delta.  The  notation $\langle \ldots \rangle$  stands for the average over independent searches on the same landscape. The results of the simulations exhibited in the next sections were obtained by  averaging over $10^5$ searches.
Notice that  $\langle t^*_L \rangle = 1/\left [ 1 - \left ( \lambda_N \right)^L \right ]$ is the expected number of trials for a group of $L$ independent agents to find the global maximum. 
In particular, for $N=12$ we have  $\lambda_{12} \approx 0.99978$ and  $\langle t^*_1 \rangle \approx  4545$. Since 
$\left ( \lambda_{12} \right)^L \approx e^{- L \left ( 1 - \lambda_{12} \right )}$ we have 
$\langle C \rangle \approx \langle t^*_1 \rangle / 2^{12} \approx 1.110$ for $L \ll \langle t^*_1 \rangle$  and
$\langle C \rangle \approx L/ 2^{12} $ for $L \gg \langle t^*_1 \rangle$.

\section{Regular lattices}\label{sec:latt}

Here we consider the case that the agents are fixed at the $L$ sites of a one-dimensional lattice with periodic boundary conditions  (i.e., a ring) and can interact with their $M/2$ left neighbors as well as with their $M/2$ right neighbors. Hence $ M \leq L-1$ is the coordination number of the lattice. This configuration is important because  one can vary the  connectivity $M$ of the lattice without the risk of  producing disconnected clusters.

%----------------------------------------------------------------------------------------------------------------
\begin{figure}
\centering
\includegraphics[width=0.48\textwidth]{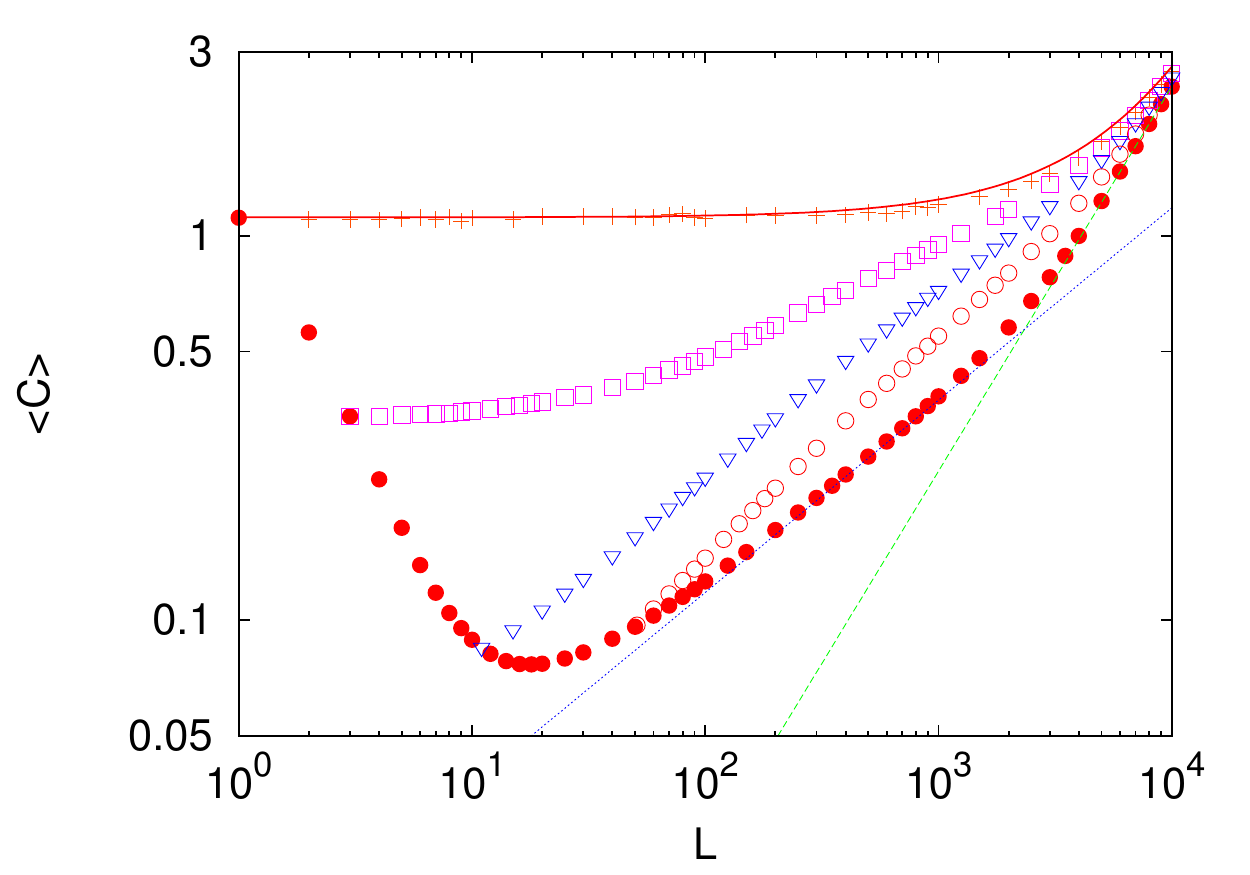}
\caption{(Color online) Mean rescaled computational cost $\langle C \rangle $ for
the imitation probability  $p=0.5$ as function of the number of sites $L$ of a regular one-dimensional lattice (ring)  with coordination number $M =2  (\square)$, $M=10 (\triangledown)$,  and $M= 50   (\Circle)$. The results for the fully connected lattice $M = L -1 (\CIRCLE) $ as well as for the independent search $(+)$ are also shown. The solid  curve is  Eq.\ (\ref{Cind}), the dashed line is the linear function $\langle C \rangle = L/2^{12}$ and the dotted line is the fitting $\langle C \rangle = 0.76 \left (L/2^{12} \right )^{1/2}$ in the range $L \in \left [10^2,10^3 \right ]$. The parameters of the NK landscape are $N=12$ and $K=0$. The landscape exhibits a single maximum.}
\label{fig:1}
\end{figure}
%------------------------------------------------------------------------------------------------------------
%

Figure \ref{fig:1} shows the performance of a group of $L$ agents at the task of finding the global optimum of a smooth landscape ($K=0$) in the case the copying and the elementary move operations are equally probable, i.e., $p=0.5$. We find that, regardless of the number of agents $L$, the best performance is achieved by the fully connected lattice $M = L -1$ and the performance degrades smoothly as the connectivity decreases. This can be explained by  the absence of local maxima in the landscape for $K=0$. Since all sequences display faithful information about the location of the global maximum, the higher the  information flow between them, the better the performance of the group.  In addition, we find that regardless of the connectivity  the computational cost decreases as the imitation probability increases. We recall that for $p \approx 1$  only the model string, which is  represented by several copies when $L$ is large,  is likely to perform the elementary move; all other strings imitate the model.  This  reduces greatly  the effective group size since the strings are concentrated in the vicinity of the model string,  which cannot accommodate more than   $L=N=12$ strings without duplication of work. This is the reason we observe the degradation of the performance when the  group size increases beyond the optimal value $L \approx N$. Notice that for $K=0$ the imitative learning search  always performs better than the independent search. 

At this stage it is prudent to point out that a constant computational cost implies that the time $t^*$ necessary to find the global maximum decreases linearly with the number of agents. On the other hand, a computational cost that grows linearly with the group size  means that adding more agents to the group does not  affect  $t^*$.  For group sizes in the range $L \in \left [10^2,10^3 \right ]$, Fig.\ \ref{fig:1} reveals a sublinear decrease of $t^*$, since  $\langle C \rangle $  increases with $L^{1/2}$ and, therefore, $t^*$ decreases with $L^{-1/2}$. According to our criterion for efficiency, as a collaborative strategy  the imitative search is efficient only in the range $L < 10$ where we find  the superlinear scaling $t^* \propto L^{-2}$.

%----------------------------------------------------------------------------------------------------------------
\begin{figure}
\centering
\includegraphics[width=0.48\textwidth]{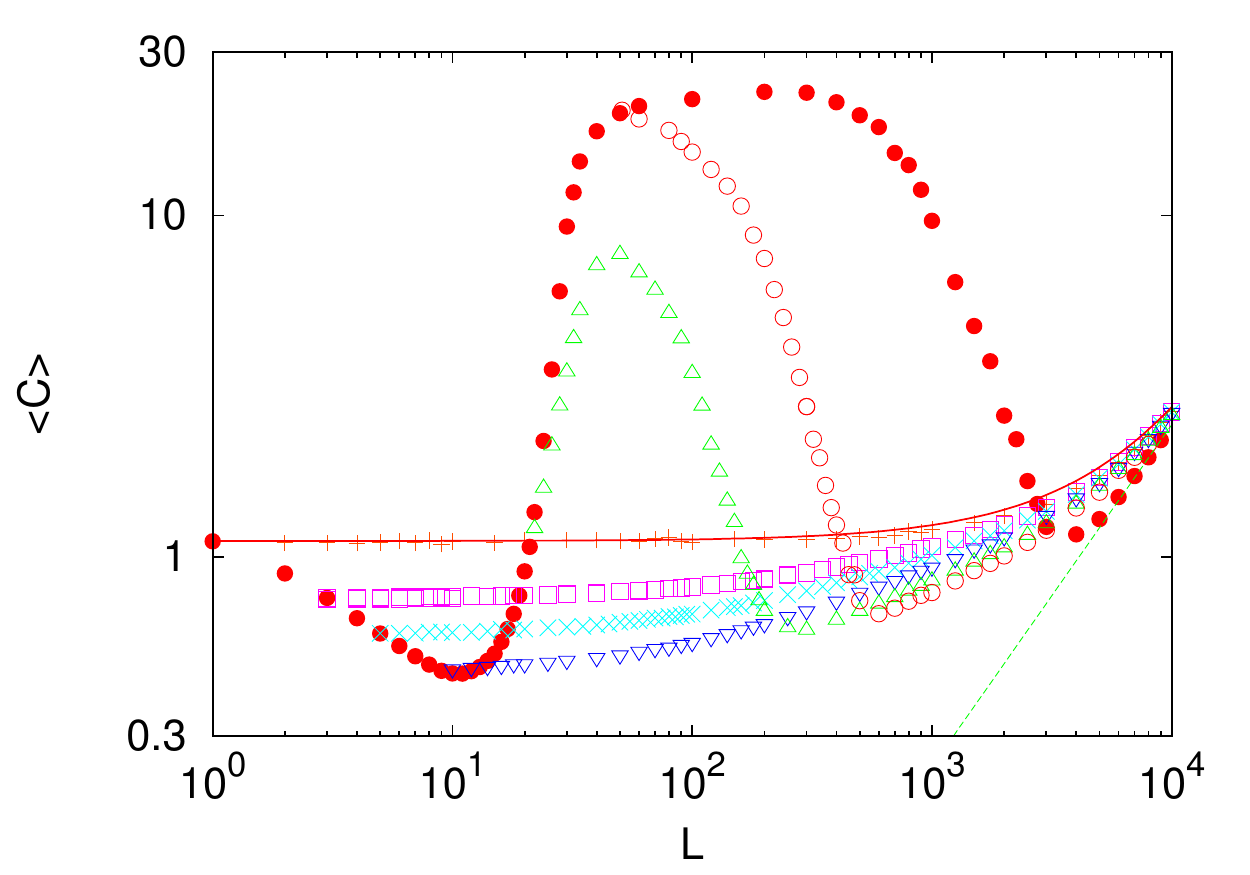}
\caption{(Color online) Mean rescaled computational cost $\langle C \rangle $ for
the imitation probability  $p=0.5$ as function of the number of sites $L$ of a regular one-dimensional lattice (ring) with coordination number $M =2  (\square)$, $M=4 (\times)$, $M=8 (\triangledown)$, $M=20  (\triangle)$ 
and $M= 50   (\Circle)$. The results for the fully connected lattice $M = L -1 (\CIRCLE) $ as well as for the independent search $(+)$ are
shown also. The solid  curve is  Eq.\ (\ref{Cind})
and the dashed line is the linear function $\langle C \rangle = L/2^{12}$. The parameters of the NK landscape are
$N=12$ and $K=4$. The landscape exhibits 52 local maxima and a single global maximum.}
\label{fig:2}
\end{figure}
%------------------------------------------------------------------------------------------------------------
%

\begin{figure}
\centering
\includegraphics[width=0.48\textwidth]{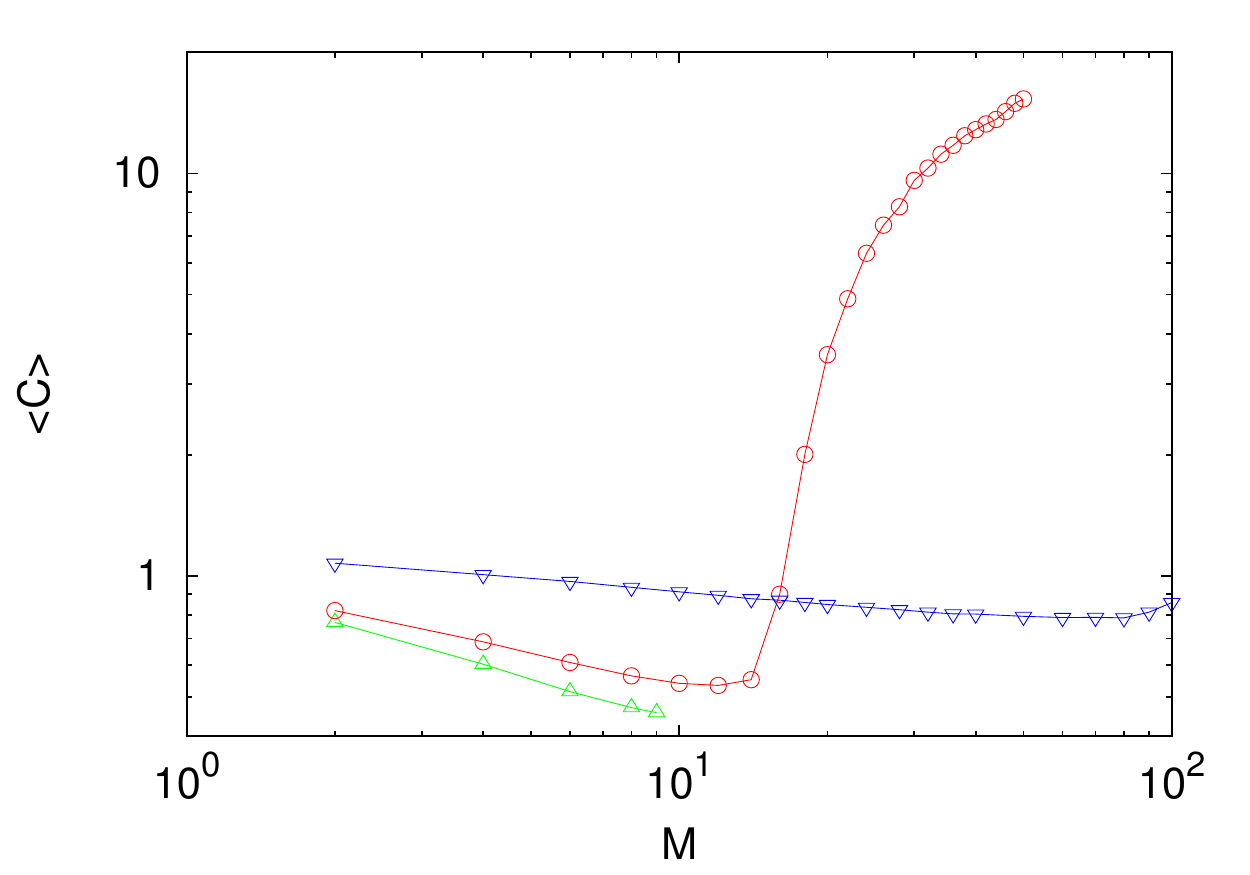}
\caption{(Color online) Mean rescaled computational cost $\langle C \rangle $ for
the imitation probability  $p=0.5$ as function of the  coordination number number  $M \leq L-1$ for  regular one-dimensional lattices (rings) with $L =10  (\triangle)$, $L=100  (\Circle)$ and $L=1000  (\triangledown)$.  The lines are guides to the eyes. The parameters of the NK landscape are
$N=12$ and $K=4$. The landscape exhibits 52 local maxima and a single global maximum.}
\label{fig:3}
\end{figure}

Figure \ref{fig:2} shows the performance for a rugged landscape ($K=4$) having 52 local maxima and a single global maximum.   Because of the presence of local maxima that may trap the sequences in their neighborhoods, the computational cost exhibits a much more complex dependence on the model parameters than in the previous case of a single-maximum landscape. This figure reveals many instructive facts about the imitative search strategy.  In particular, this strategy may be disastrous, in the sense that it performs much worse than the independent search, for large groups  with  high connectivity (i.e., $L \in \left [ 20,2000 \right]$ and $M > 15$ for the data of Fig. \ref{fig:2}). In this case, the decrease of the connectivity  and consequently of the rate of information flow through the network greatly improves the performance of the imitative search. Most interestingly, however, is the finding that for small groups (say, $L < 10$) the fully connected system is the optimal one, as shown in Fig.\  \ref{fig:3}. The same conclusion holds  true in the case the number of agents is greater than the size of the solution space ($L >  2^{12}$) since it is very likely that some string will be close to the global maximum and it is thus advantageous to disseminate this information quickly through the group. For groups of intermediate size  (say, $L=100$) the detrimental effects of increasing the connectivity 
of the lattice are startling as illustrated in Fig.\ \ref{fig:3}.  

%
%----------------------------------------------------------------------------------------------------------------
%------------------------------------------------------------------------------------------------------------
%

\begin{figure}
\centering
\includegraphics[width=0.48\textwidth]{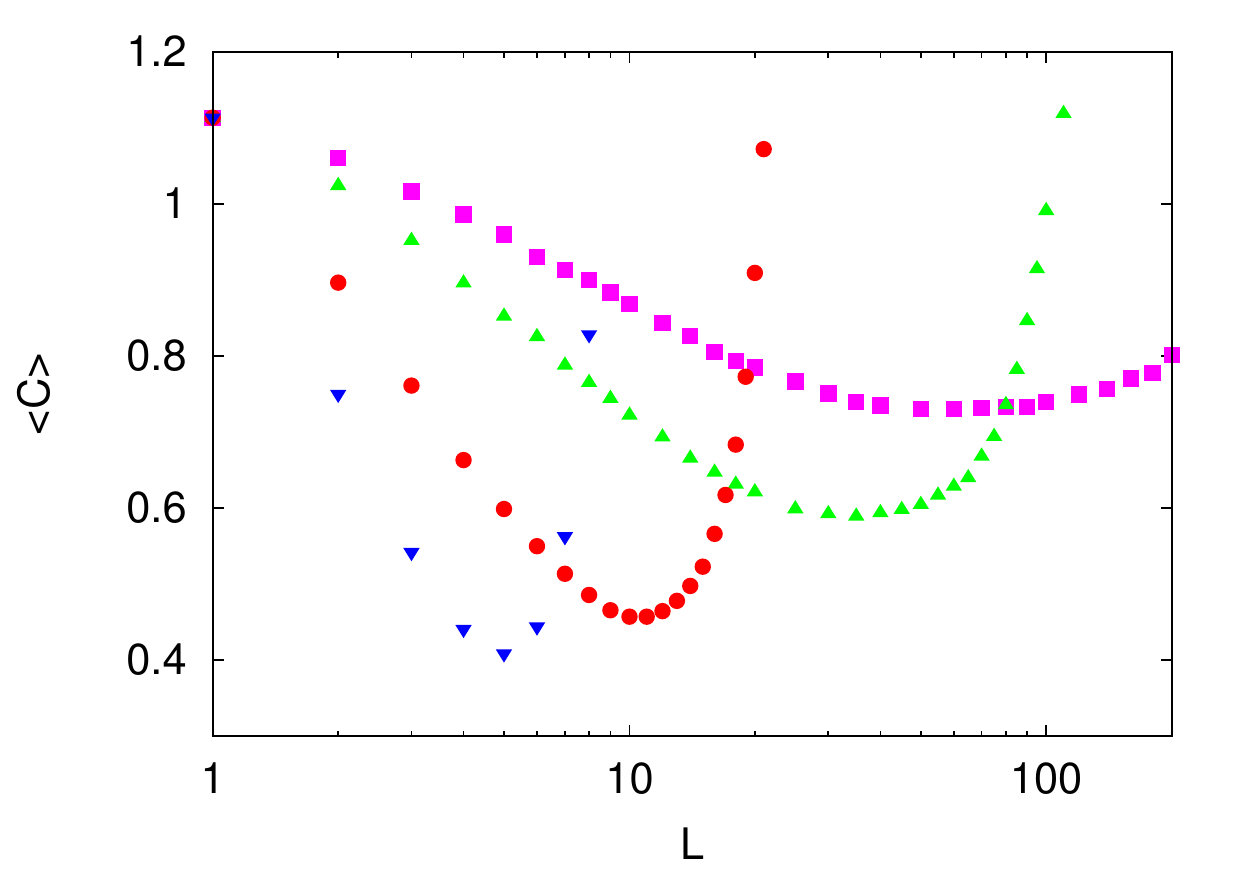}
\caption{(Color online) Mean rescaled computational cost $\langle C \rangle$ for the fully connected lattice as function of the number of sites $L$ for
the imitation probability $p=0.2 (\blacksquare)$, $p=0.3 (\blacktriangle)$, $p=0.5 (\CIRCLE)$ and $p=0.7 (\blacktriangledown)$. The parameters of the NK landscape are
$N=12$ and $K=4$. The landscape exhibits 52 local maxima and a single global maximum.}
\label{fig:new}
\end{figure}

The main conclusion  of this analysis is that the optimal setting for the imitative search is a small group of size $L^*$,
which depends on the imitation probability $p$ as well as on details of the landscape, where the agents are fully connected.  In particular,
$L^*$ decreases with increasing $p$ as illustrated in Fig.\ \ref{fig:new}: for $p >0.98$ we find $L^* = 2$ whereas for $p \to 0$ we find
$L^* \approx 2^N$. Although the decrease of  the connectivity between  agents can boost  significantly the performance of the group  in the regime  $ L^* < L \ll 2^N$, the leading factor for optimality is the group size,  with the lattice connectivity playing a coadjuvant role (see  Fig.\ \ref{fig:3}).

%----------------------------------------------------------------------------------------------------------------
%------------------------------------------------------------------------------------------------------------
%

\section{Complex networks}\label{sec:comp}

The previous section focused on the influence of varying the connectivity of the regular network on the problem-solving performance of the group. Here we consider
networks with the same average connectivity but very different organizations, as reflected by the degree distribution -- the degree $k_i$ of a node $i$ is given by its number of connections. In particular, we consider  scale-free networks  generated by the Barab\'asi-Albert algorithm \cite{Barabasi_99}, a star topology and a regular chain with nearest-neighbors links. In the three topologies there are $L$ nodes and $L-1$ links. In addition, to examine the influence of long-range interactions we consider small-world networks  \cite{Watts_98} as well as a network exhibiting community structure \cite{Girvan_02}.

\subsection{Scale-free networks}

A scale-free network is a network whose degree distribution follows a power law when the number of nodes $L$ is very large. For our purposes, the interesting feature of this  topology, besides being a good approximation to social and biological networks \cite{Albert_02}, is that it introduces a distinction between agents, since the nodes of the network can exhibit  very different degrees.  Here we use the  Barab\'asi-Albert algorithm for generating random scale-free networks which  exhibit  the degree distribution 
$P \left ( k \right ) \sim k^{-3}$ in the asymptotic limit $L \to \infty$ \cite{Barabasi_99}. This algorithm is based on a preferential attachment mechanism and network growth. Thus, at each time step, a new node $i$ with $m_0$ connections is added to the network. The probability that a node $j$, which is present in the network, will receive a connection from $i$ is proportional to $k_j$ (the degree of $j$), i.e., $P(i\to j) = k_j/\sum_l k_l$. In the case  we start with a single node and $m_0=1$, the resulting network will exhibit  $L-1$ connections at the $L$th step. In addition, if $m_0=1$ then the network  will show a tree-like structure~\cite{Albert_02}.

%----------------------------------------------------------------------------------------------------------------
\begin{figure}
\centering
\includegraphics[width=0.48\textwidth]{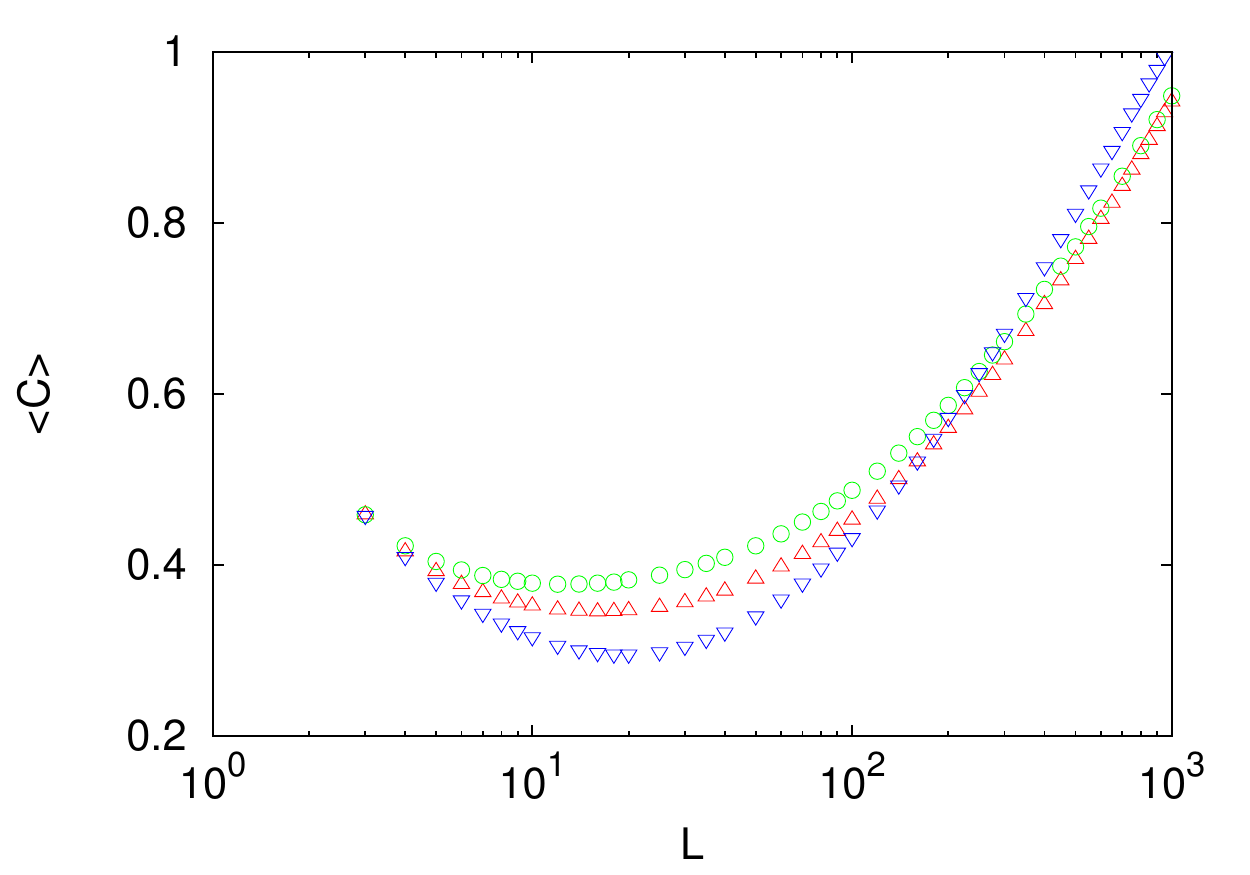}
\caption{(Color online) Mean rescaled computational cost $\langle C \rangle $ for
the imitation probability  $p=0.5$ as function of the number of sites $L$ for a chain with coordination number $M =2$  $(\Circle)$,
scale-free networks generated according to the Barab\'asi-Albert algorithm with initial degree $m_0=1$ $(\triangle)$, and a star topology $(\triangledown)$.
In each run we used a different random scale-free network.
The parameters of the NK landscape are
$N=12$ and $K=0$. }
\label{fig:4}
\end{figure}
%------------------------------------------------------------------------------------------------------------
%

\begin{figure}
\centering
\includegraphics[width=0.48\textwidth]{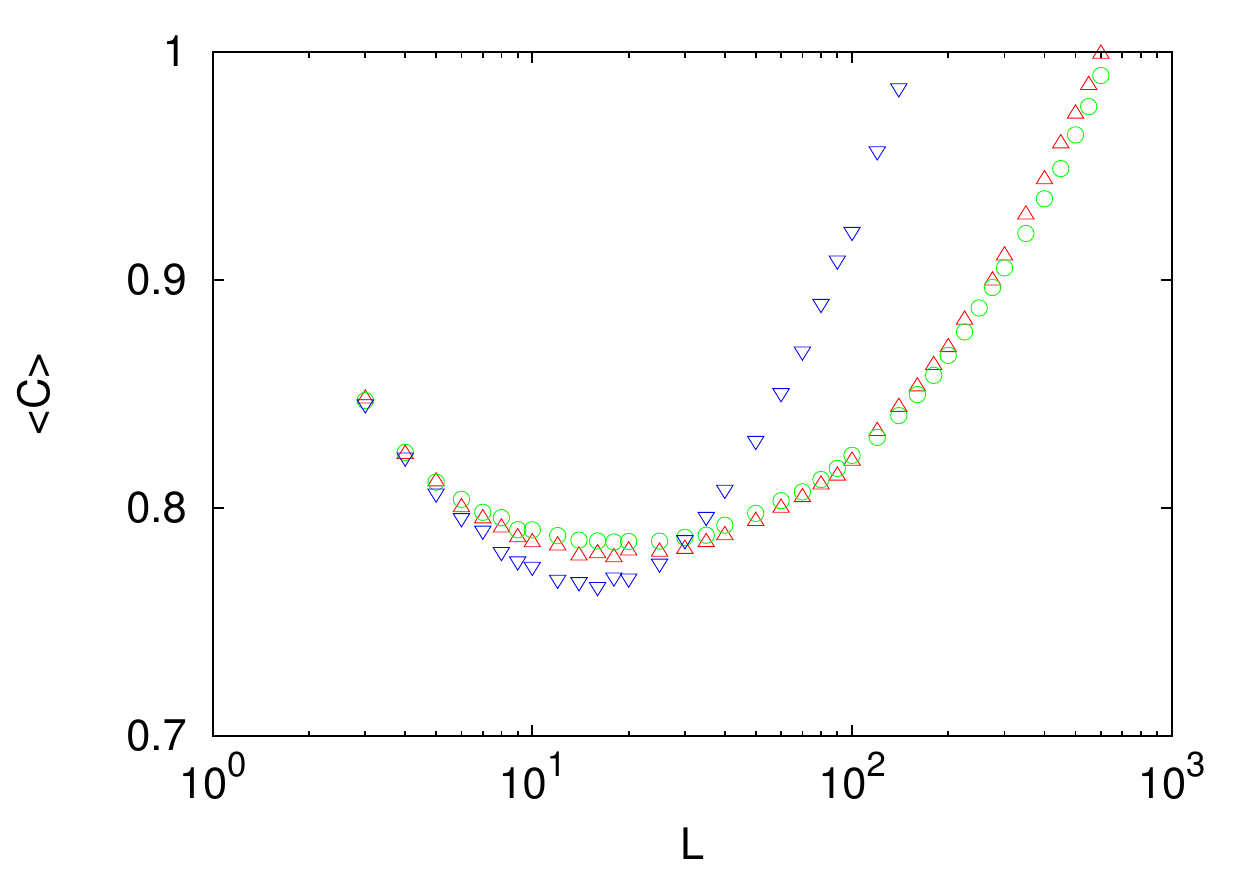}
\caption{(Color online) Same as Fig.\ \ref{fig:4}  but for a rugged NK landscape with parameters 
$N=12$ and $K=4$. }
\label{fig:5}
\end{figure}

Notice that the size of the neighbors group of each node (i.e., the influence network) varies in scale-free networks, contrariwise to the regular networks considered in the last section. The results of the imitative search in a smooth landscape  ($K=0$) and in a rugged landscape ($K=4$) are summarized in Figs.\ \ref{fig:4} and \ref{fig:5}, respectively, in the case the $L$ agents are located at the nodes of a scale-free network. For the sake  of comparison, we consider also a star topology where  $L-1$ agents are connected to a central agent (a super-spreader), and a regular one-dimensional lattice with free boundary conditions and coordination number  $M=2$. In the three network topologies considered in these figures, the number of agents is $L$ and the number of links is $L-1$. In addition, for $L=3$ the three topologies produce identical graphs.

Since in the absence of local maxima ($K=0$) the model string always
displays faithful information about the location of the global maximum, the faster this information spreads among the agents, the better the performance of the group. Figure \ref{fig:4} indicates that for $L$ not too large, the star topology maximizes the information flow among agents and so it yields  the best performance among the three topologies.  The chain topology minimizes the information flow and so it yields the worst performance, whereas the scale-free topology exhibits a performance  level that is intermediate to those two extremes.  However, the star topology has obvious drawbacks that become apparent for large $L$ only
($L > 200$ for the data of Fig.\  \ref{fig:4}): in the case the fittest agent is not the central agent,  the useful, new  information it carries is not available for imitation to the other agents, who can imitate the central agent only.  Actually, that piece of information can even be lost if the fittest agent is chosen as
the target agent before it is imitated by the central one. 

In the presence of local maxima ($K=4$), we find a similar pattern, as shown in Fig.\ \ref{fig:5}, but the failure of the star topology  already for $L>30$ is probably due to the spreading of inaccurate information about the global optimum by the central agent. In fact, once the central agent reaches a local maxima it quickly attracts the rest of the group towards it.

We may interpret the central agent of the star topology as a type of blackboard, which the other agents must access to get hints about the location of the global maximum \cite{Englemore_88}. In fact, when the central agent is selected  as the target and the imitation operation is  chosen, it copies the best performing agent of the previous trial and this information become available to the other agents, which can then access the blackboard with probability $p$.  Given
the severe bottleneck to the information flow in the star topology, it comes to a surprise that it works so well for small groups.

%
%----------------------------------------------------------------------------------------------------------------
%------------------------------------------------------------------------------------------------------------
%

%
%----------------------------------------------------------------------------------------------------------------
\begin{figure}
\centering
\includegraphics[width=0.48\textwidth]{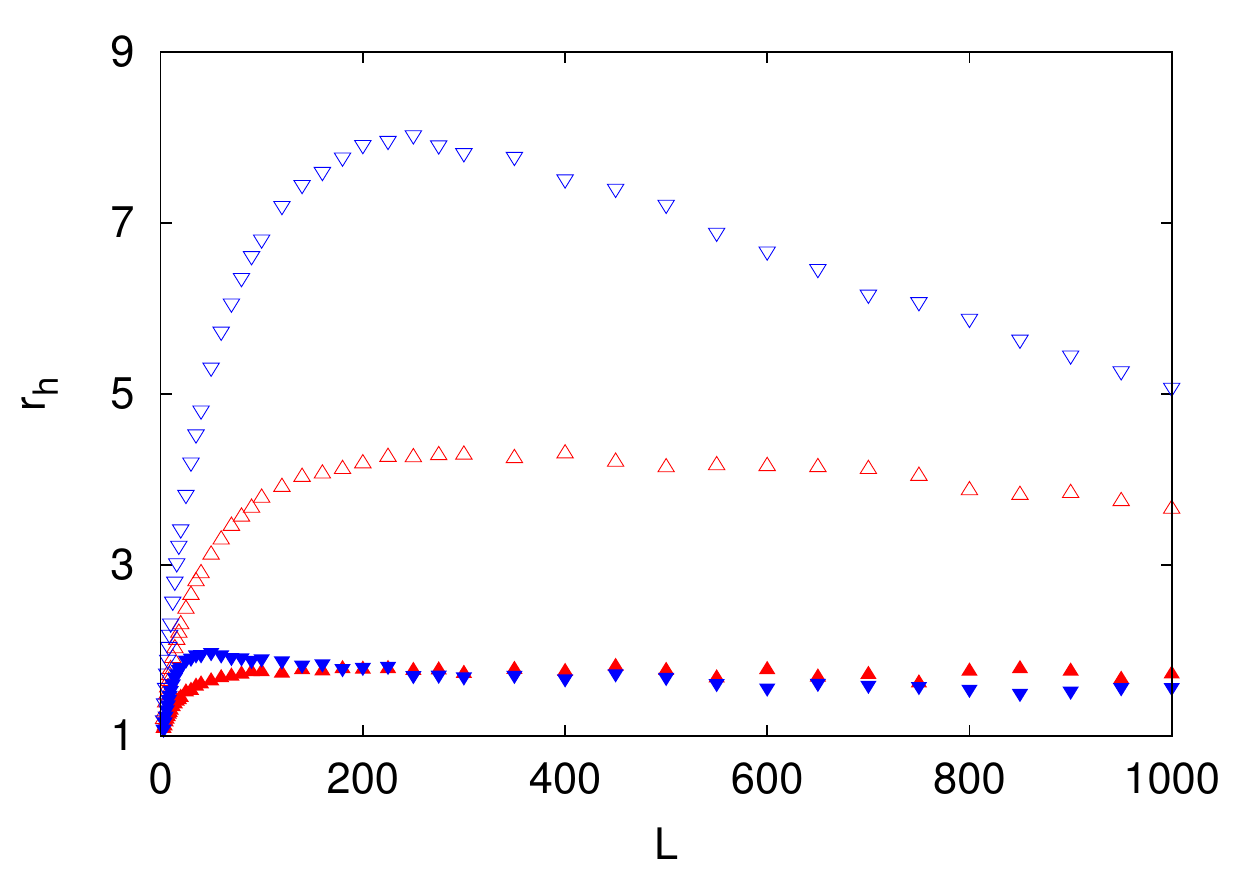}
\caption{(Color online) Ratio $r_h$ between the probability $P_h$  that the agent with the highest degree in the network finds the global maximum 
and the baseline probability $1/L$ for the case the $L$ agents are equiprobable to find that maximum. The symbols  $(\triangle)$ are the results for
scale-free networks generated according to the Barab\'asi-Albert algorithm, whereas the symbols $(\triangledown)$  are for the star topology.
In each run we used a different random scale-free network. The imitation probability is  $p=0.5$ and the parameters of the NK landscape are
$N=12$, $K=0$ (open symbols) and $K=4$ (filled symbols). }
\label{fig:6}
\end{figure}
%------------------------------------------------------------------------------------------------------------
%

For  topologies where the degree of the nodes (or agents) is not constant, we can estimate the probability $P_h$ that the  agent with the highest degree is the one that finds the global maximum.   Since in the case all agents have the same probability of finding the global maximum (e.g., for the regular lattices discussed in Section \ref{sec:latt}) this probability is $1/L$, it is convenient to consider the  ratio  $r_h = P_h/\left ( 1/L \right ) = L P_h \geq 1$ which  gives a measure of the influence of the degree of an agent on its chances of hitting that maximum.
In  Fig.\ \ref{fig:6} we show  $r_h$ as function of the group size for the star and the scale-free topologies. In the limit of very large $L$, chances are that the global maximum already shows up in the setting of the initial population, so we have $r_h \to 1$ in this limit. The influence of the degree on the ratio $r_h$ depends on the ruggedness of the landscape: it is significant for smooth landscapes but  marginal for rugged landscapes. It is interesting that for the star topology the regime where  $r_h$ decreases with increasing $L$ roughly coincides with the regime where the star topology  performs less well than the other
topologies considered in Figs.\ \ref{fig:4} and \ref{fig:5}.

%
%----------------------------------------------------------------------------------------------------------------
\begin{figure}
\centering
\includegraphics[width=1\columnwidth]{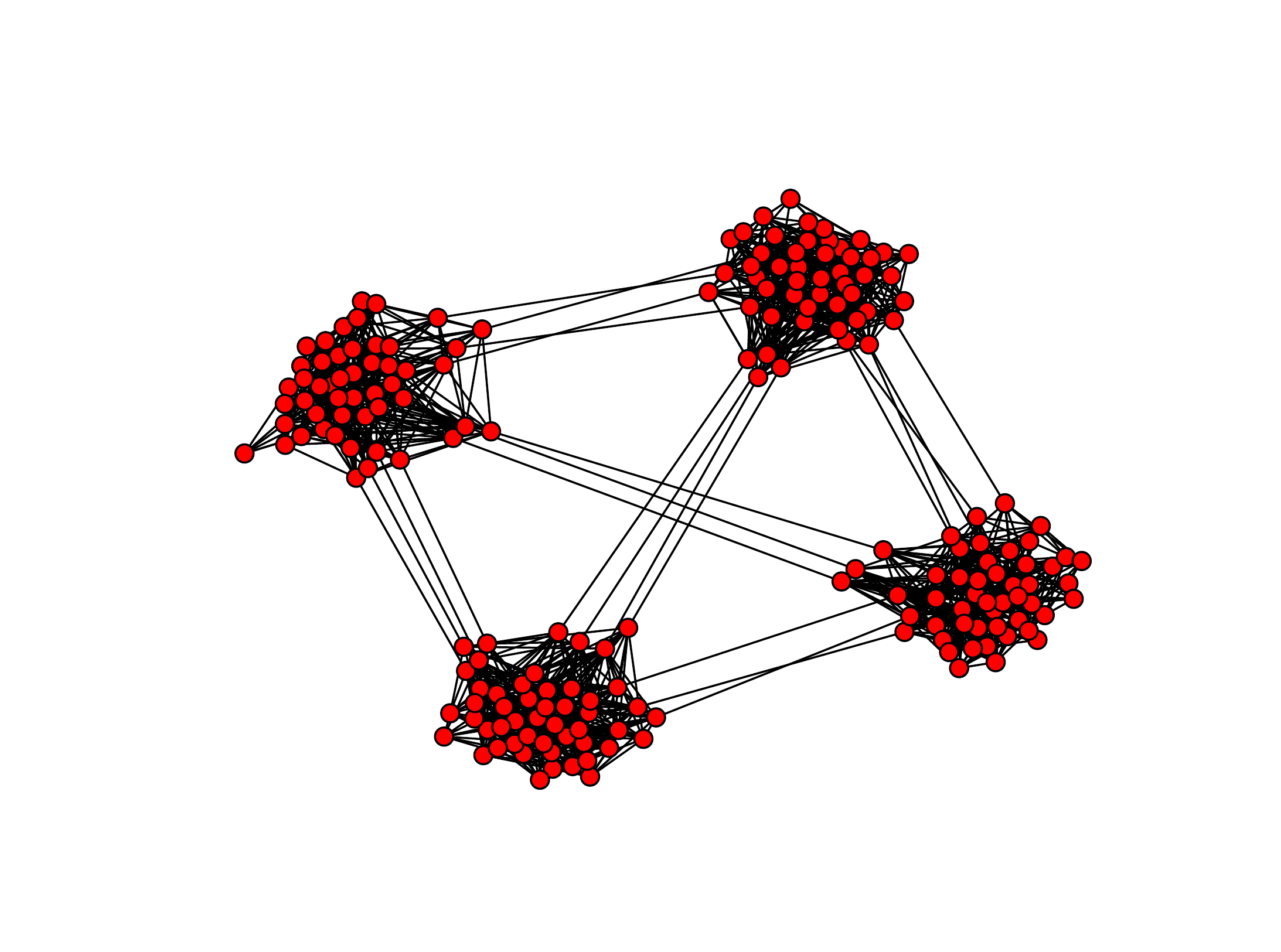}
\caption{(Color online) Network with a community structure consisting of $L=200$ nodes divided in 4 groups of 50 nodes each. The average connectivity of the network is
approximately $15.2$.   }
\label{fig:7}
\end{figure}
%------------------------------------------------------------------------------------------------------------
%

\subsection{Community structure and small-world  networks}

To explore  the effects of the structure of the population on the performance of the imitative search,
we consider a network with $L=200$ agents divided in 4 clusters of $50$ agents each  as illustrated in Fig.\ \ref{fig:7}. The probability that two nodes of the same cluster are connected is $0.3$, whereas nodes in different clusters are connected according to a probability of $0.001$. The total number of links is 1519 so the average connectivity of the network is approximately $15.2$.  We consider, in addition,  small-world networks which allow the interpolation between  structureless (random) graphs and regular lattices  \cite{Watts_98}. 

In particular, the small-world networks we consider here are generated using the Watts and Strogatz algorithm. Explicitly, we begin  with  a  one-dimensional lattice of $L$ nodes, each node connected to $M$ neighbors ($M/2$ on each side), and periodic boundary conditions.  
Then  for every node  $i= 1,\dots, L$  we rewire the links  between $i$ and $j=i+1$ (the sums are done modulo $L$) with probability $\beta$. Rewiring of a link is done by replacing the original neighbor of node $i$ by a random node chosen uniformly among  all possible nodes that avoid self-loops  and link duplication. This procedure is then repeated for the links between $i$ and $j=i+2$, $i$ and $j=i+3$,  and so on until the links between $i$ and $j=i + M/2$. The case
$\beta=1$ corresponds to random graphs with average connectivity $M$ and the case $\beta =0$ to a regular ring with number of coordination $M$. 

\begin{figure}
\centering
\includegraphics[width=0.48\textwidth]{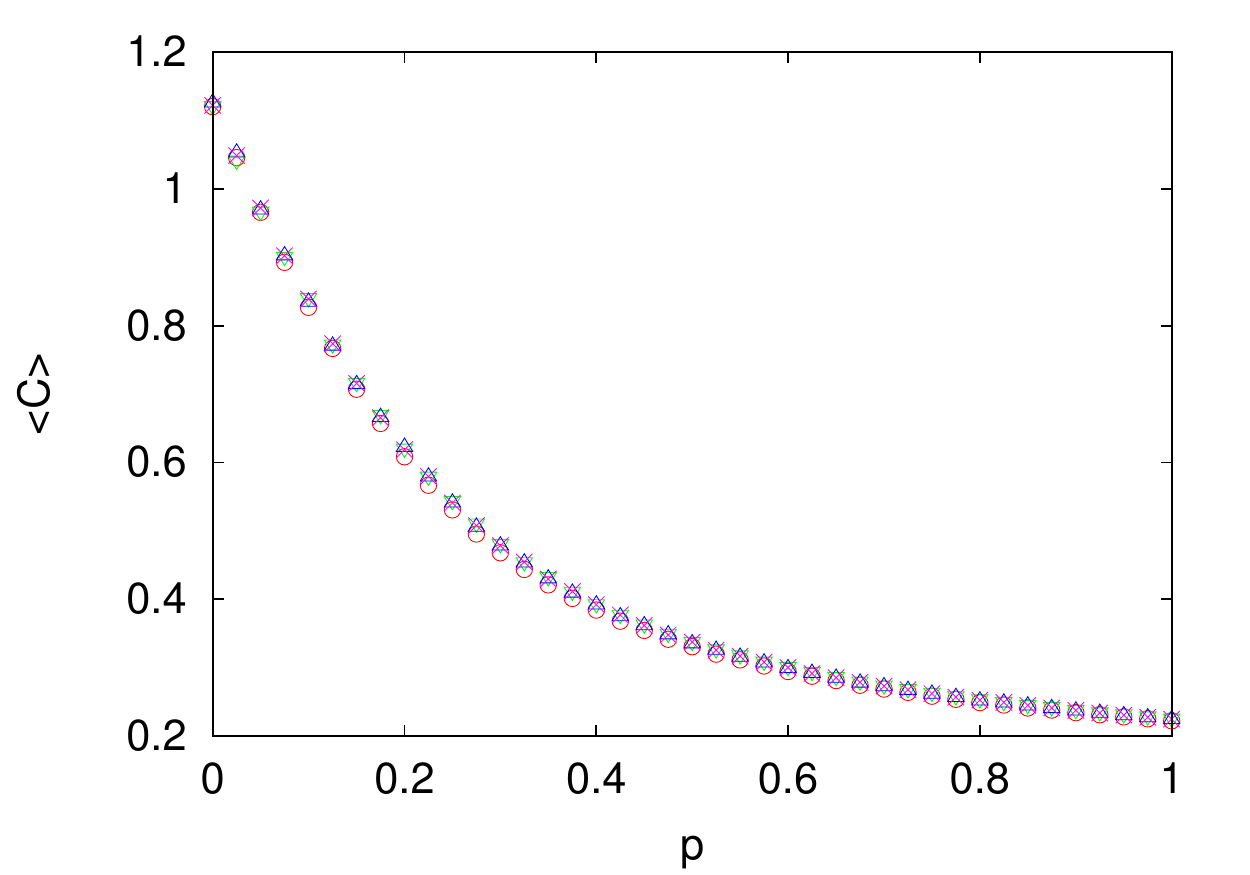}
\caption{(Color online) Mean rescaled computational cost $\langle C \rangle $ as function of the imitation probability  $p$ for the  network with community structure
illustrated in Fig.\ \ref{fig:7}  $(\Circle)$, and for small-world networks with average connectivity $14$ and rewiring probability $\beta = 1$ $(\triangle)$,  $\beta = 0.2$ $(\times)$, and  $\beta = 0$ $(\triangledown)$.
The number of nodes is $L=200$ for the three networks and 
the parameters of the   smooth NK landscape are
$N=12$ and $K=0$.   }
\label{fig:8}
\end{figure}

\begin{figure}
\centering
\includegraphics[width=0.48\textwidth]{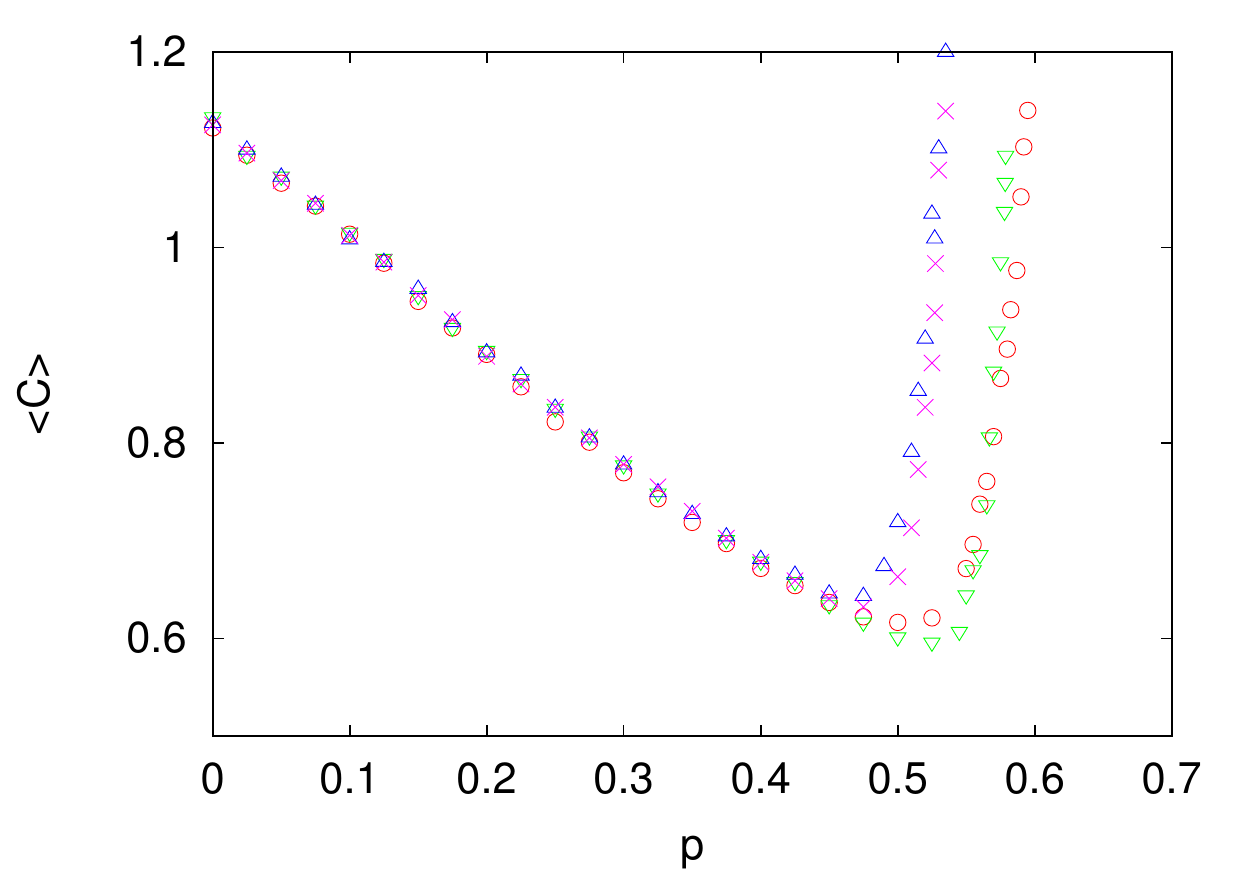}
\caption{(Color online) Same as Fig.\ \ref{fig:8} but for the rugged NK landscape
of parameters
$N=12$ and $K=4$.   }
\label{fig:9}
\end{figure}

Figures  \ref{fig:8} and \ref{fig:9} show the mean computational cost for small-world networks with rewiring probability $\beta =0, 0.2$ and  $1$, and for
the network with community structure  shown in Fig.\ \ref{fig:7}. We find that  $\langle C \rangle $ is completely insensitive to variations on the structure of the network for  smooth landscapes (see Fig.\ \ref{fig:8}). This conclusion applies to rugged landscapes as well, except for $p$ close to the  boundary of the region where the imitative search fails to find the global maximum  (see Fig.\ \ref{fig:9}). In this case the sensitivity is extreme. For instance,  for $p=0.58$   searches using random  networks ($\beta =1$) fail to find the global maximum, whereas searches using regular ($\beta =0$) or community networks succeed. In fact, for large values of the imitation probability,  the risk of the search  being stuck in  local maxima is very high and so long-range links which speed up the flow of information through the network may become notably detrimental to the   imitative search performance. However, in the absence of
local maxima (Fig.\ \ref{fig:8}) the best performance is attained by  imitating the model string and allowing only their clones to explore the landscape through the elementary move.

 %
%----------------------------------------------------------------------------------------------------------------
%------------------------------------------------------------------------------------------------------------
%

  %
%----------------------------------------------------------------------------------------------------------------
%------------------------------------------------------------------------------------------------------------
%

\begin{figure}
\centering
\includegraphics[width=0.48\textwidth]{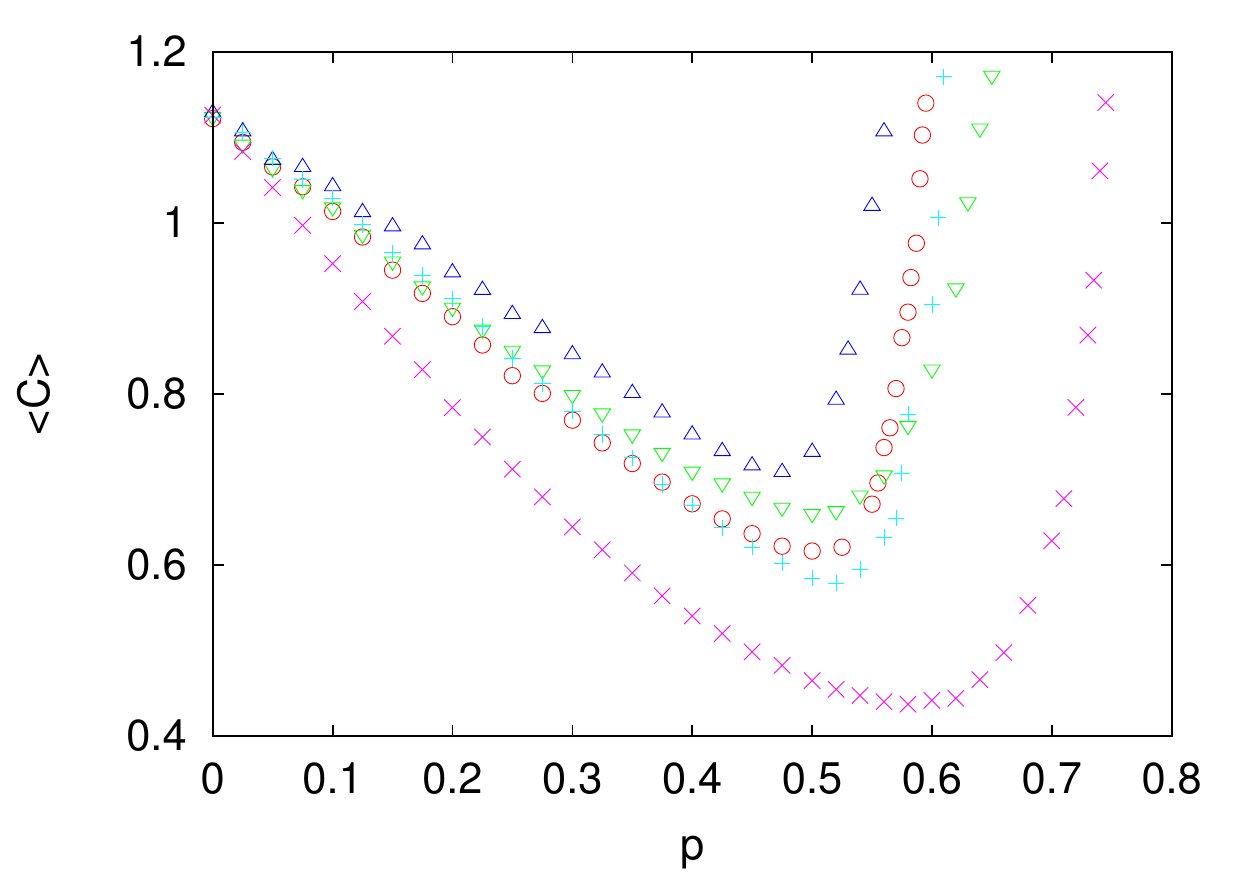}
\caption{(Color online) Mean rescaled computational cost $\langle C \rangle $  as function of   the probability of imitation $p$ 
for the  network with community structure illustrated in Fig.\ \ref{fig:7} and five realizations (different symbols) of the NK fitness landscape  with $N=12$ and $K=4$.   }
\label{fig:10}
\end{figure}

To check whether  the specific realization of the NK rugged fitness landscape we used in our study has unduly influenced our conclusions,   we have considered  four random realizations of the landscape with $N=12$ and $K=4$ in addition to that realization.  We note that for smooth landscapes ($K=0$) all realizations are equivalent. The comparison between the mean computational costs for  those five realizations   is shown in Fig.\ \ref{fig:10}. The results are qualitatively the same. The quantitative differences are due to variations in the number of local maxima as well as in their distances to the global maximum for the different landscape realizations.   It is reassuring to note
that the initial decrease of  the mean cost with  the imitation probability $p$, which indicates that (moderate) imitation is a sensible cooperation strategy,  as well as the existence of a threshold imitation value beyond which the group fails to find the global maximum, are robust  properties of the imitative learning search on rugged landscapes.

%----------------------------------------------------------------------------------------------------------------
%------------------------------------------------------------------------------------------------------------
%

 \section{Discussion}\label{sec:conc}
 
The tenet of our approach to  study cooperative problem-solving systems is  that  group size and organization are selected so as to maximize group-level performance as measured  by  the number of trials necessary to find the global maxima of fitness landscapes. Differently from the definition of network optimality  in biology, where there is a direct selection pressure to reduce  the number of connections between entities because of their  building and maintenance costs \cite{Clune_13}, here we assume that the connections between agents are costless. As a result, we find that the   optimal group organization is obtained by a  relatively small group of  fully connected agents. The optimal group size $L^*$ is a decreasing function of  the imitation probability $p$ (see Fig.\ \ref{fig:new})  and is roughly proportional to the logarithm of the size of the solution  space \cite{Fontanari_15}.
In our model, the problem-solving efficiency decreases with the group size for $L > L^*$   because of the duplication of work resulting from the decrease of the
diversity of the group members  due to  imitation. More traditional selection pressures that contribute to limit the size of  groups are resource competition,  parasite transmission and the  managing of the social relationships \cite{Kurvers_14,Dunbar_92}.

The manner  the information on the location of the global maximum spreads among the agents, which is determined by 
the network structure,  as well as the accuracy of that information can greatly affect the  problem-solving performance of the group. For instance,
we find that for smooth landscapes the decrease of the agents' connectivity  is always 
harmful  to the performance, regardless of the group size $L$.  For rugged landscapes, however, this decrease
can be highly beneficial for large  groups (i.e., $L > L^*$). In fact, because the model agent may broadcast misleading information in this case,
it is advantageous to slow down the information transmission so as to allow the agents more time to explore the solution space away from the 
neighborhoods of the local maxima. The same conclusion holds in the case of centralization:  the presence of super-spreaders,  which enhance the diffusion of information, is beneficial provided the information is accurate.  Decentralized networks  are less susceptible to the diffusion of inaccurate information.
It is interesting that for small groups (i.e., $L < L^*$) both centralization and
full connectivity are beneficial, regardless of the accuracy of the information. In addition, the presence of long-range links has no effect on the performance of the group, except for a very narrow region of the model parameters. Notice that the evolution of other dynamical processes, such as synchronization or cascade failures, are strongly affected by the network structure~\cite{Barthelemy_08}. Therefore, the weak influence of long-range links on the performance of the imitative learning search  is not an expected result. 

An interesting question that we can address within our minimal model framework  is whether the fixed topology  of a social network can  enhance the fitness of some group members \cite{Waters_12}. In this case the accuracy of the transmitted information seems to be the crucial ingredient. In fact, for smooth landscapes we find that  highly-connected agents are very likely to be the ones that find the global maximum, but for rugged landscapes the fitness advantage of  the  central agents   is marginal only (see Fig.\ \ref{fig:6}). Another interesting
issue, which we plan to address in a future work, is the study of   adaptive networks,  in which the network  topology  itself changes during the  search  \cite{Gross_08,Perra_12}. The difficulty here is to find a suitable procedure to update the topology, since simply adding more connection to the model (fittest) agent can be catastrophic if that agent is near a local maximum.

Finally, we note that our study departs from the vast literature on the game  theoretical approach to the evolution of cooperation  \cite{Axelrod_84} where  it is usually taken for granted  that  mutual cooperation is beneficial to the players. Here we consider a problem solving scenario  and a specific cooperation mechanism (imitation) aiming at determining in which conditions cooperation is beneficial. 
The  findings that in some cases
cooperation results in  maladaptive behavior  \cite{Laland_98,Laland_11}  and that for very large groups efficiency is achieved by minimizing communication  \cite{Moffett_11} support our conjecture that  the efficacy of imitative learning could influence  the  size and organization of the groups of social animals.

\acknowledgments
The research of JFF was partially supported by grant
2013/17131-0, S\~ao Paulo Research Foundation
(FAPESP) and by grant 303979/2013-5, Conselho Nacional de Desenvolvimento 
Cient\'{\i}\-fi\-co e Tecnol\'ogico (CNPq). FAR acknowledges CNPq (grant 305940/2010-4) and FAPESP (grant 2013/26416-9) for financial support.

\end{document}